\documentclass[twocolumn,aps,prl,showpacs,raggedbottom,nobalancelastpage,superscriptaddress]{revtex4-1}

\usepackage{graphicx}
\usepackage{amsmath}
\usepackage{amsfonts}

\begin{document}
\title{Quantum chaos and effective thermalization} 

\author{Alexander Altland}

\affiliation{Institut f{\"u}r Theoretische Physik,
Universit{\"a}t zu K{\"o}ln, K{\"o}ln, 50937, Germany}

\author{Fritz Haake}

\affiliation{Fakult\"at f\"ur Physik, Universit\"at Duisburg-Essen, 47048 Duisburg, Germany}

\begin{abstract}
  We demonstrate effective equilibration for unitary quantum dynamics
  under conditions of classical chaos. Focusing on the paradigmatic
  example of the Dicke model, we show how a constructive description of the
  thermalization process is facilitated by the Glauber $Q$ or Husimi
  function, for which the evolution equation turns out to be of Fokker-Planck
  type. The equation describes a competition of classical drift and quantum
  diffusion in contractive and expansive directions. By this mechanism
  the system follows a 'quantum smoothened' approach to equilibrium,
  which avoids the notorious singularities inherent to classical
  chaotic flows. 
\end{abstract}
\pacs{}
\maketitle
How do thermally isolated many particle quantum systems relax towards
stationary states? Along with the recent advances in quantum optics
and cold atom experimentation, this fundamental question of
theoretical physics presents itself under new perspectives. The
interacting, and in many instances thermally decoupled many particle
states realizable in cold atom devices show novel forms of relaxation
which can be monitored at hitherto unimaginable degrees of
resolution. Examples of unconventional relaxation phenomena include
the formation of negative temperature states\cite{Mosk:2005fk}, the
buildup of textures in Brillouin zones by an interplay of interaction
and single particle kinematics\footnote{U. Schneider \textit{et al.},
  arXiv:1005.3545v1}, or the relaxation into ergodic and other types
of 'non-thermal' distributions~\cite{Rigol:2008ys,Goldstein:2010fk,Polkovnikov:2011zr}.

There appears to be a concensus that for fully
developed chaos, and in the absence of conservation laws, 'relaxation'
towards equipartition on a shell of conserved energy ('microcanonical
distribution') represents a universal paradigm. Recent work describes
the ensuing equilibrated state in terms of
energy eigenfunctions whose phase space profiles are presumed
ergodic~\cite{Rigol:2008ys,Goldstein:2010fk,Polkovnikov:2011zr}. While 
computationally efficient, this scheme does not reveal the 
physical processes leading to thermalization.

In this letter we approach the phenomenon from a more dynamical
perspective.  We present our case for the paradigmatic example of the
Dicke model~\cite{Dicke:1954kx} which describes the coupling of a
(large) spin and an oscillator.  At a critical coupling strength, the
model undergoes a quantum phase transition into a 'superradiant'
phase, characterized by a nonvanishing mean oscillator amplitude and
chaotic dynamics \cite{Emary:2003qf}. In view of the recent
experimental observation of that phase
transition~\cite{Baumann:2010vn}, the issue of thermalization has
become concrete.

Our approach hinges on a manageable description of the impact of
quantum fluctuations on the semiclassical dynamics.  A representation
of the quantum density operator, by Glauber's $Q$-function (aka Husimi
function), turns out to be the key.  Von Neumann's equation for the
density operator assumes the form of a Fokker Planck equation for $Q$.
The equilibration processes can be then understood by appreciating the
competition of classical drift and quantum diffusion in expansive and
contractive directions of the 'flow' supporting $Q$.  We augment the
general discussion by an exact solution of a toy model which arguably
reflects the essence of the full problem.

In the end we will indicate that our methodology is applicable to a
whole class of chaotic dynamics.  Even certain periodically kicked
systems qualify.

\textit{Dicke Model ---.} We write the Dicke Hamiltonian as
\begin{equation}
  \label{H}
  \hat{H}=\hbar\Big\{\omega_0 \hat J_z+\omega a^\dagger a +
               g\textstyle{\sqrt{\frac{2}{j}}}(a+a^\dagger)\hat J_x\Big\}\,.
\end{equation} 
Here, $\hat J_a, a=x,y,z$ are spin operators acting in a spin-$j$
representation, and $a/a^\dagger$ are photon annihilation/creation
operators. The first two terms in \eqref{H} respectively describe spin
precession about the $J_z$-axis with frequency $\omega_0$ and harmonic
oscillation with frequency $\omega$. The last term describes the
coupling of spin and oscillator with coupling constant $g$. Crucially,
the interaction contains the so called antiresonant terms
$J_+a^\dagger +J_-a$ where $J_\pm=J_x\pm i J_y$ are the familiar
raising and lowering operators. This fact makes the model distinct
from an integrable variant where these terms are neglected (rotating
wave approximation). At a critical value of the coupling strength,
$g_c\equiv \sqrt{\omega\omega_0}/2$, the non-integrable model
\eqref{H} undergoes a quantum phase transition into a 'superradiant
phase'. For values $g>g_c$, the photon 'amplitude' $a$ builds up a
nonvanishing excitation value, and the dynamics becomes globally
chaotic.

\textit{Coherent state representation ---.} In view of the largeness
of the spin, $j\gg 1$, we find it convenient to represent the theory in
terms of coherent states~\cite{Arecchi:1972kx,Glauber:1976uq,Scully:1994ly},
\begin{align}
  \label{eq:1}
  |z\rangle \equiv {1\over (1+|z|^2)^j} e^{ z\hat J_-}\, |j,j\rangle,
\end{align}
where $z\in\Bbb{C}$ and $|j,j\rangle$ is a 'maximum-weight' eigenstate
of $J_z$, i.e.  $J_z |j,j\rangle=j |j,j\rangle$. The states
$|z\rangle$ are unit normalized, $\langle z|z\rangle=1$. They entail
the expectation values $ \langle z|\hat J_a|z\rangle =j l_a$, where
$l_{x,y,z}$ are the three components of a unit vector
$\mathbf{l}=(\sin\theta \cos\phi,\sin\theta\sin\phi,\cos\theta)^T$,
whose angular orientation is defined through $ z=e^{i\phi}
\tan(\theta/2)$. Each $|z\rangle$ has minimum angular uncertainty,
characterized by the solid angle $4\pi/(2j+1)$ which defines a Planck
cell on the unit sphere.  The overcomplete set
$\{|z\rangle\}$ provides a resolution of unity as
$\openone=\frac{2j+1}{\pi }\int\frac{d^2z}{(1+zz^*)^2} |z\rangle \langle
z|$.

Similarly, oscillator coherent states are defined
as~\cite{Glauber:2007fk}
\begin{equation}
  \label{cohstateosc}
  |\alpha\rangle=e^{-\alpha\alpha^*/2}\,e^{\alpha a^\dagger}|0\rangle,
\end{equation}
where $\alpha\in\Bbb{C}$ and $|0\rangle$ is the vacuum,
$a|0\rangle=0$. The states $|\alpha\rangle$ assign a minimal
uncertainty product to displacement and momentum such that these
quantities are 'confined' to a single Planck cell located at $x\propto
(\alpha+\alpha^*)\sqrt{\hbar},\,p\propto
i(\alpha-\alpha^*)\sqrt{\hbar}$ within the classical phase space. The
completeness relation here reads $\openone=\frac{1}{\pi}\int\,
d\alpha\, |\alpha\rangle \langle \alpha|$.

We next represent the system's time dependent density operator $\hat
\rho(t)$ in terms of a coherent state based quasiprobability
density. Among the many possible choices the Glauber $Q$-function,
\begin{equation}
  \label{Q}
  Q(\alpha,z)=\frac{2j+1}{\pi (1+zz^*)^2}\langle \alpha,z|\rho|\alpha,z\rangle\,,
\end{equation}
turns out to be the most convenient one by far, for our purposes. We
here denote by $|\alpha,z\rangle\equiv |\alpha\rangle
\otimes|z\rangle$ the overall coherent state for spin and
oscillator. Using the completeness relations given above one sees that
$Q$ is normalized as $\int d\alpha dz\, Q(\alpha,z)=1$ and yields
expectation values of (anti-normal ordered) operators as
 $ \langle a^m a^{\dagger n}\rangle=
                \int d\alpha dz\, \alpha^m{\alpha^*}^{n}\,Q(\alpha,z)$\,.
By its definition, $Q$ exists and is non-negative, $Q(\alpha,z)\geq
0$, for any density operator $\rho$. The latter property allows $Q$ to
converge to the classical phase space density as $\hbar\to 0$.  As the
most rewarding property we shall presently find that $Q$ enables us to
map the quantum evolution equation $ i\hbar d_t\hat \rho = [\hat
H,\hat \rho]$ onto a differential equation of Fokker-Planck type,
optimally suited to describe the dissipation-free equilibration.

\begin{figure*}
  \centering
  \includegraphics[width=18cm]{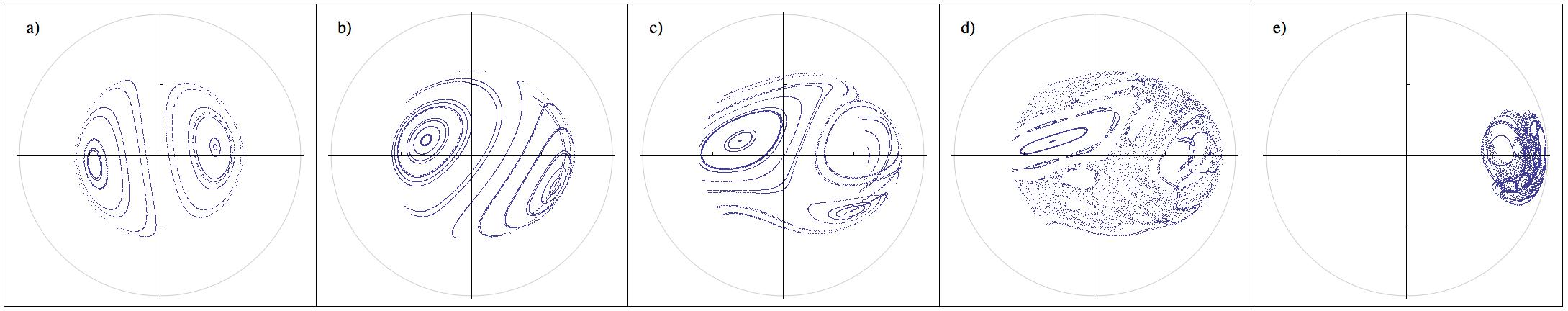}
  \includegraphics[width=18cm]{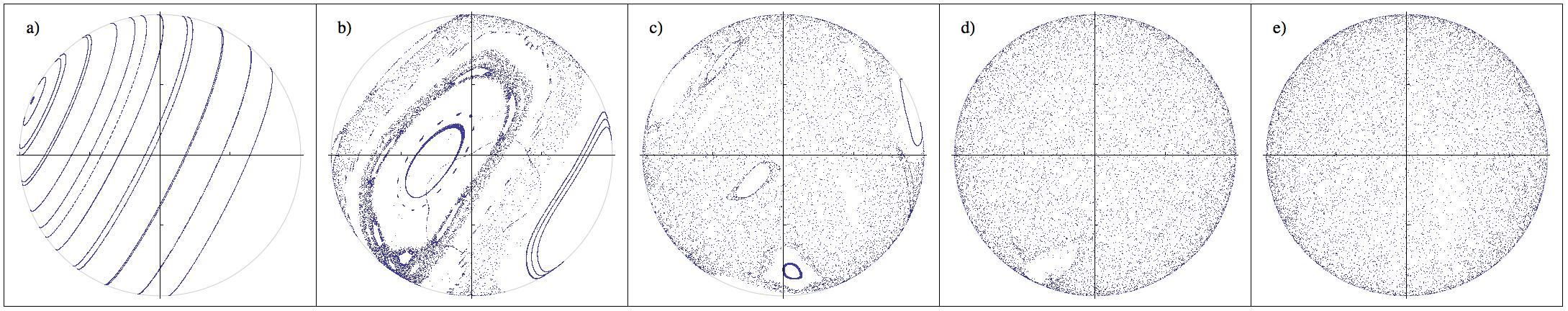}
  \caption{Poincar\'e sections generated by monitoring the projection
    $(l_x,l_y)$ of $\mathrm{l}$ in the southern hemisphere at fixed
    values of the phase $\psi$. For each parameter value, $g/g_c$,
    nine trajectories of different on-shell initial conditions are
    sampled. Upper row: energy $\Delta \epsilon\simeq 0.2|\epsilon_0|$
    above the ground state and values $g/g_c$ a) $0.2$, b) $0.7$, c)
    $0.9$, d) $1.01$, e) $1.5$. Lower row: energy $\Delta
    \epsilon\simeq 20|\epsilon_0|$.}
   \label{fig:poincarelow}
  \end{figure*}

  \textit{Quantum dynamics ---.} Using identities such as $\hat J_x
  |z\rangle \langle z| = \big((1-z^2)\partial_z +2 j {z+z^\ast\over
    1+|z|^2}\big)|z\rangle \langle z|$ and $a^\dagger |\alpha\rangle
  \langle \alpha|=(\partial_\alpha + \alpha^\ast)|\alpha\rangle
  \langle \alpha|$,
  it is straightforward to show that the evolution equation of the
  distribution function $Q$ assumes the Fokker-Planck form
\begin{widetext}
  \begin{align}
  \label{quantumevol}
  \dot{Q}&=(\mathcal{L}+\mathcal{L}_{\rm diff})Q\,\\ \nonumber
  \mathcal{L} &=i\partial_\alpha\left(\omega\alpha + 
                g\sqrt{2j}\frac{z+z^*}{1+|z|^2} \right)
          +i\partial_z\left(\!\!-\omega_0 z+
                 \frac{g}{\sqrt{2j}}(1-z^2)(\alpha+\alpha^*)\!\right)
               +{\rm c.c.}\,,
            \qquad   
   \mathcal{L}_{\rm diff} = \frac{ig}{\sqrt{2j}}\partial_{\alpha}\partial_z  (1-z^2)
                     +{\rm c.c.}\,.
\end{align}
\end{widetext}
Most remarkably, derivatives terminate at second order. Less prudent
choices of a quasiprobability might have brought the plague of higher
derivatives \footnote{For instance, the frequently employed Wigner
  function $W$ has evolution equations with only odd-order derivatives
  such that quantum features start with third order. One then
  sometimes dispatches all but the first-order derivative terms
  ('truncated Wigner approximation'), thence relegating the effects of
  quantum mechanics to the uncertainty carried by the initial
  distribution. Interestingly, for chaotic dynamics $W$ typically
  develops substructures within Planck cells\cite{Zurek:2001uq}; such
  structures cannot show up in $Q$ which arises from $W$ by averaging
  over, roughly, a Planck cell.}.

To understand the meaning of the first order, or drift term, it is
convenient to introduce canonically conjugate action angle variables
$(I,\psi)$ as $\alpha \equiv \sqrt{jI } \exp(i\psi)$ for the
oscillator and the canonical pair $(\cos\theta,\phi)$ for the
spin. 
It is then a straightforward (if tedious) matter to show that the
drift operator becomes the classical Liouville operator,
${\cal L} Q=\{h,Q\}$, where the effective Hamilton function
\begin{align}
\label{eq:2}
  h=\omega_0 \cos\theta + \omega I + g \sqrt{8I}
    \cos\psi \sin\theta \cos\phi,
\end{align}
obtains from the operator (\ref{H}) by substituting
$\hat{\mathbf{J}}\to j \mathbf{l}$, $a\to \sqrt{jI} \exp(i\psi)$ and
dividing out $\hbar j =L$. The operator ${\cal L}$ thus describes the
drift of the quasi-probability along the classical
trajectories of the Hamiltonian flow.

The distinctive feature of the classical dynamics is global chaos in
the superradiant regime, $g>g_c$~\cite{Emary:2003qf}. Referring for an
in-depth discussion to \footnote{A. Altland and F. Haake, to be
  published}, we here merely note that for $g>g_c$ and for excitations
energies only slightly above the ground state bound $E_0\simeq
-j\omega_0\big({g\over g_c}\big)^2$ ergodic dynamics is observed. This
is illustrated in the Poincar\'e sections of
Fig.~\ref{fig:poincarelow}, where the top/bottom row are for low
($\Delta E/|E_0|\simeq 0.20$)/high ($\Delta E/|E_0|\simeq 30$)
excitation.  Both sections (e) reflect dominance of chaos and
indicate equilibration.  In particular, the
low-excitation portrait (e) reveals chaos as dominant, even though the
energy shell includes but a small part of the Bloch
sphere.

\textit{Quantum diffusion ---.} How does quantum mechanics interfere
with the classical drift?  Quantum mechanics enters the Fokker-Planck
equation (\ref{quantumevol}) through the diffusion operator ${\cal
  L}_\mathrm{diff}$. A key feature of the diffusion operator is the
absence of second order derivatives w.r.t. only oscillator
($x_{1,2}=I,\psi$) or spin ($x_{3,4}=\cos\theta,\phi$), phase space
variables. The 'diffusion matrix' $D$ defining the operator ${\cal
  L}_\mathrm{diff}\equiv \sum_{ij}\partial_{x_i}\partial_{x_j}
D_{ij}(x)$ thus possesses a block off-diagonal \textit{chiral}
structure, $D= \left(\begin{smallmatrix} 0&d\cr d^\dagger &0
\end{smallmatrix}\right)$. While there will be no need to spell out the explicit
dependence of the chiral block $d$ on the variables
$x=(I,\psi,\cos\theta,\phi)$, we emphasize its smallness in
$j^{-1}$. The 'chiral' block structure of $D$ entails a secular
equation for the eigenvalues of the form $\lambda^4-\lambda^2\mbox{
  tr}\,dd^\dagger+\det{dd^\dagger}=0$. The four eigenvalues of $D$
thus come in two pairs $\pm\sqrt{a},\pm\sqrt{b}$ where $a,b$ are the
eigenvalues of the non-negative $2\times 2$ matrix $dd^\dagger$. 
Each eigenvalue is associated with an eigenvector corresponding to
diffusive spreading $(+)$ or 'anti-diffusive' shrinking $(-)$. 
These four 'quantum' directions form a Cartesian frame wherein the
four directions distinguished by the chaotic drift lie askew --- one
expansive (unstable), one contractive (stable), and two neutral (along
the flow and transverse to the energy shell).

Since $Q$ is guaranteed existence and positivity, quantum diffusion
(along the eigenvectors of $D$ with positive eigenvalues) must oppose
the contraction along the classically stable manifold, thus preventing
the buildup of singular phase space structures.  Deterministic
contraction and quantum diffusion balance at quantum scales
$1/\sqrt{j}\sim\sqrt{\hbar}$. At shorter 'length' scales (which can be
brought into play by initial states squeezed along the classically
stable direction), the second order differential diffusion operator
becomes dominant (as can be seen by straightforward scaling
estimates); it prevents further contraction and smoothes the
distribution.  By contrast, at the \textit{large} scales generated in
the classically unstable direction, the quantum (anti-)diffusive
correction is of no significance. Rather, expansion continues
uninhibited, accompanied by the chaotic 'folding' necessitated by the
compactness of the energy shell.

\textit{Toy model ---.} It is instructive to illustrate the interplay
of classical drift and quantum diffusion on an analytically solvable
toy model, provided by the Fokker-Planck equation
$\dot{P}=\big(\partial_p p -\partial_x
x+D(\partial_p^2-\partial_x^2)\big)P$; the diffusion constant must be
imagined small, $0<D\ll1$ and of quantum origin, $D\propto \hbar$.
Anti-diffusive contraction and deterministic stretching are carried by
$x$ according to $\mbox{var}_t(x)=\big(\mbox
{var}_0(x)-D\big)e^{2t}+D$, where the variance is calculated using an
averaging prescription $\langle f(x,p)\rangle =\int dx dp
f(x,p)P(x,p)$. We infer that the minimum allowable 'quantum' scale of
$x$ is $\sqrt{D}$. If $\mbox {var}_t(x)$ is of the order but larger
than $D$, the characteristic scale for (the variance of) $x$ will grow
to ('macroscopic') order unity within the Ehrenfest time
$\sim\ln(1/D)$.  On the other hand, diffusive expansion and
deterministic contraction are carried by $p$ as $\mbox{var}_t(p)=
\big(\mbox{var}_0(p) -D\big)e^{-2t}+D$. The balancing scale for $p$ is
$\sqrt{D}$. If the initial variance of $p$ is much larger than $D$
(say, of the 'macroscopic' order unity), the exponential decrease to
the order of $D$ takes a time of the order $\ln (1/D)$, again the
'Ehrenfest time'.

Generalizing the toy model one may
choose the axes of classical contraction and expansion skew to the
principal axes of quantum (anti)diffusion. Positivity of all variances
then restricts the relative orientations: clearly,
the classically stable direction must not coincide with the axis of
antidiffusion, or else the respective variance would go negative
within a finite time. Referring for a detailed discussion of the
ensuing 
correlations in the Dicke system to~\cite{Note3}, we here merely note that its
guaranteed existence protects the $Q$-function from such type of
alignment, 
save for exceptional regions in phase space.

\textit{Thermalization ---.} Turning back to the full problem, we
conclude that the interplay of deterministic expansion/contraction,
quantum diffusion, and chaotic folding (the latter of course absent in
the toy model) will spread out any initial distribution over the
compact energy shell.  Location, $E$, and width, $\Delta E$ of the
shell are determined by the initial state, with ${\Delta E \over E }\sim
j^{-1/2}$ a minimal value for coherent state initial
distributions. Given any fixed phase space resolution, $\Delta x$, and
a characteristic width $x_0\gg \Delta x$ of the initial distribution,
the quantum and a purely classical description of the flow, resp.,
both predict full coverage of the energy shell for time scales
$t>\tau\ln(\Delta x_0/\Delta x)$, where $\tau$ is the Lyapunov time.
However, important differences occur in the dynamical buildup of the
equilibrated state: while the classical approach describes
equilibration in terms of the formation of an infinitely filigree
structure of alternating high and vanishing phase space density, ---
the result of continued stretching and folding of the initial
distribution ---, the quantum distribution remains smooth (on scales
$\sim j^{-1/2}$) all the way along. This hallmark of quantum chaotic
propagation should be experimentally visible by current spectroscopic
techniques\cite{Chaudhury:2009fk}. Finally, let us remark on the
universality of the above picture with regard to variations in the
initial conditions. Coherent states, being isotropically 'supported'
by a single Plack cell, come closest to the classical fiction of a
single phase space point. They naturally qualify as initial states for
the equilibration described above. Equally well suited are squeezed
minimum uncertainty states, even if squeezed along the classically
stable direction; the diffusion then smears the pertinent uncertainty
to the characteristic quantum scale mentioned. Yet broader states 
succumb to equilibration even more willingly.

\textit{Beyond the Dicke model ---.} While the relevance of 'quantum
diffusion' operators to the description of the long time dynamics in
chaotic quantum systems has been noted before, previous
work~\cite{AlLa1} has added these contributions by hand. Our present
analysis exemplifies how quantum diffusion emerges naturally.
Evolution equations with derivatives terminating at second order are
not an exclusive privilege of the Dicke model. Rather, whenever a
chaotic dynamical system has a Hamiltonian of the form of a
second-order polynomial in the pertinent observables and allows for a
coherent-state-based $Q$-function, we expect a Fokker-Planck equation
to govern the time evolution of $Q$ and everything to go through in
much the same way as above.  Examples are autonomous $SU(3)$
dynamics~\cite{Gnutzmann:2000uq} whose Hamiltonians contain terms of
first and second order in the  $SU(3)$ generators. Certain kicked
systems qualify as well. Most notable among those is the kicked
top \cite{Haake} whose near classical quantum behavior has recently
been observed experimentally~\cite{Chaudhury:2009fk}. Genuine
many-body systems also have  Fokker-Planck
equations for $Q$, provided the Hamiltonians are quartic in creation
and annihilation operators but contain no antiresonant
terms. Examples of much current interest are Bose-Hubbard systems
\cite{Trotzky:kx,PhysRevA.79.021608}. Finally, chaotic dynamics where the
evolution equations of $Q$ contains higher-order derivatives can
allow for reasonable Fokker-Planck approximations, if good arguments
are available for neglecting third and higher derivatives. We shall
cover such extensions of our work in a separate  publication \cite{Note3}.

Summarizing, we have discussed the principles whereby the
probability flux of a classically chaotic quantum systems covers its
shell of conserved energy. The dynamics crucially hinges on
effectively diffusive quantum propagation
interfering with the deterministic exponential contraction inherent to
classically chaotic flows. (In the expansive directions of the flow,
quantum corrections are negligible.) The net effect of this
competition is the smoothing of a flow which in a classical system
would soon turn into a singular structure. We discussed these phenomena on
the example of the Dicke model where the (Husimi representation of
the) quantum flow equation assumes
the particularly handy form of a Fokker-Planck equation. While the
Fokker-Planck representation pertains to other  quantum
systems, 
it is far from generic. In general, one will meet higher-order,
differential equations. However, power
counting arguments~\cite{Note3} suggest that at the microscopic length scales $\sim
\hbar^{1/2}$ where quantum diffusion becomes effective, higher terms
in the expansion will be sub-dominant. 
It thus
stands to reason that the smoothing of the classical flow by
mechanisms similar to those discussed above is a general effect.

We gratefully acknowledge helpful discussions with T. Brandes,
P. Braun, C. Emary, T. Esslinger, V. Gurarie, J. Larson,
M. Lewenstein, and M. Ku\'s. Work supported by the
SFB/TR12 of the Deutsche Forschungsgemeinschaft.

\bibliography{~/Documents/REVTEX/BIBLIOGRAPHY/my}

\begin{thebibliography}{10}%
\makeatletter
\providecommand \@ifxundefined [1]{%
 \ifx #1\undefined \expandafter \@firstoftwo
 \else \expandafter \@secondoftwo
\fi
}%
\providecommand \@ifnum [1]{%
 \ifnum #1\expandafter \@firstoftwo
 \else \expandafter \@secondoftwo
\fi
}%
\providecommand \enquote [1]{``#1''}%
\providecommand \bibnamefont  [1]{#1}%
\providecommand \bibfnamefont [1]{#1}%
\providecommand \citenamefont [1]{#1}%
\providecommand\href[0]{\@sanitize\@href}%
\providecommand\@href[1]{\endgroup\@@startlink{#1}\endgroup\@@href}%
\providecommand\@@href[1]{#1\@@endlink}%
\providecommand \@sanitize [0]{\begingroup\catcode`\&12\catcode`\#12\relax}%
\@ifxundefined \pdfoutput {\@firstoftwo}{%
 \@ifnum{\z@=\pdfoutput}{\@firstoftwo}{\@secondoftwo}%
}{%
 \providecommand\@@startlink[1]{\leavevmode\special{html:<a href="#1">}}%
 \providecommand\@@endlink[0]{\special{html:</a>}}%
}{%
 \providecommand\@@startlink[1]{%
  \leavevmode
  \pdfstartlink
   attr{/Border[0 0 1 ]/H/I/C[0 1 1]}%
   user{/Subtype/Link/A<</Type/Action/S/URI/URI(#1)>>}%
  \relax
 }%
 \providecommand\@@endlink[0]{\pdfendlink}%
}%
\providecommand \url  [0]{\begingroup\@sanitize \@url }%
\providecommand \@url [1]{\endgroup\@href {#1}{\urlprefix}}%
\providecommand \urlprefix [0]{URL }%
\providecommand \Eprint[0]{\href }%
\@ifxundefined \urlstyle {%
  \providecommand \doi [1]{doi:\discretionary{}{}{}#1}%
}{%
  \providecommand \doi [0]{doi:\discretionary{}{}{}\begingroup
  \urlstyle{rm}\Url }%
}%
\providecommand \doibase [0]{http://dx.doi.org/}%
\providecommand \Doi[1]{\href{\doibase#1}}%
\providecommand \bibAnnote [3]{%
  \BibitemShut{#1}%
  \begin{quotation}\noindent
    \textsc{Key:}\ #2\\\textsc{Annotation:}\ #3%
  \end{quotation}%
}%
\providecommand \bibAnnoteFile [2]{%
  \IfFileExists{#2}{\bibAnnote {#1} {#2} {\input{#2}}}{}%
}%
\providecommand \typeout [0]{\immediate \write \m@ne }%
\providecommand \selectlanguage [0]{\@gobble}%
\providecommand \bibinfo [0]{\@secondoftwo}%
\providecommand \bibfield [0]{\@secondoftwo}%
\providecommand \translation [1]{[#1]}%
\providecommand \BibitemOpen[0]{}%
\providecommand \bibitemStop [0]{}%
\providecommand \bibitemNoStop [0]{.\EOS\space}%
\providecommand \EOS [0]{\spacefactor3000\relax}%
\providecommand \BibitemShut [1]{\csname bibitem#1\endcsname}%
\bibitem{Mosk:2005fk}%
  \BibitemOpen
  \bibfield{author}{%
  \bibinfo {author} {\bibfnamefont{A.~P.}\ \bibnamefont{Mosk}},\ }%
  \bibfield{journal}{%
  \Doi{10.1103/PhysRevLett.95.040403}{\bibinfo {journal} {Phys. Rev. Lett.}}\
  }%
  \textbf{\bibinfo {volume} {95}},\ \bibinfo {pages} {040403} (\bibinfo {year}
  {2005})%
  \bibAnnoteFile{NoStop}{Mosk:2005fk}%
\bibitem{Note1}%
  \BibitemOpen
  \bibinfo {note} {U. Schneider \protect \textit {et al.}, arXiv:1005.3545v1}%
  \bibAnnoteFile{NoStop}{Note1}%
\bibitem{Rigol:2008ys}%
  \BibitemOpen
  \bibfield{author}{%
  \bibinfo {author} {\bibfnamefont{M.}~\bibnamefont{Rigol}}, \bibinfo {author}
  {\bibfnamefont{V.}~\bibnamefont{Dunjko}},\ and\ \bibinfo {author}
  {\bibfnamefont{M.}~\bibnamefont{Olshanii}},\ }%
  \bibfield{journal}{%
  \bibinfo {journal} {Nature}\ }%
  \textbf{\bibinfo {volume} {452}},\ \bibinfo {pages} {854} (\bibinfo {year}
  {2008})%
  \bibAnnoteFile{NoStop}{Rigol:2008ys}%
\bibitem{Goldstein:2010fk}%
  \BibitemOpen
  \bibfield{author}{%
  \bibinfo {author} {\bibfnamefont{S.}~\bibnamefont{Goldstein}}, \bibinfo
  {author} {\bibfnamefont{J.~L.}\ \bibnamefont{Lebowitz}}, \bibinfo {author}
  {\bibfnamefont{C.}~\bibnamefont{Mastrodonato}}, \bibinfo {author}
  {\bibfnamefont{R.}~\bibnamefont{Tumulka}},\ and\ \bibinfo {author}
  {\bibfnamefont{N.}~\bibnamefont{Zanghi}},\ }%
  \bibfield{journal}{%
  \bibinfo {journal} {Phys. Rev. E}\ }%
  \textbf{\bibinfo {volume} {81}},\ \bibinfo {pages} {011109} (\bibinfo {year}
  {2010})%
  \bibAnnoteFile{NoStop}{Goldstein:2010fk}%
\bibitem{Polkovnikov:2011zr}%
  \BibitemOpen
  \bibfield{author}{%
  \bibinfo {author} {\bibfnamefont{A.}~\bibnamefont{Polkovnikov}}, \bibinfo
  {author} {\bibfnamefont{K.}~\bibnamefont{Sengupta}}, \bibinfo {author}
  {\bibfnamefont{A.}~\bibnamefont{Silva}},\ and\ \bibinfo {author}
  {\bibfnamefont{M.}~\bibnamefont{Vengalattore}},\ }%
  \bibfield{journal}{%
  \bibinfo {journal} {Rev. Mod. Phys.}\ }%
  \textbf{\bibinfo {volume} {83}},\ \bibinfo {pages} {863} (\bibinfo {year}
  {2011})%
  \bibAnnoteFile{NoStop}{Polkovnikov:2011zr}%
\bibitem{Dicke:1954kx}%
  \BibitemOpen
  \bibfield{author}{%
  \bibinfo {author} {\bibfnamefont{R.~H.}\ \bibnamefont{Dicke}},\ }%
  \bibfield{journal}{%
  \bibinfo {journal} {Phys. Rev.}\ }%
  \textbf{\bibinfo {volume} {93}},\ \bibinfo {pages} {99} (\bibinfo {year}
  {1954})%
  \bibAnnoteFile{NoStop}{Dicke:1954kx}%
\bibitem{Emary:2003qf}%
  \BibitemOpen
  \bibfield{author}{%
  \bibinfo {author} {\bibfnamefont{C.}~\bibnamefont{Emary}}\ and\ \bibinfo
  {author} {\bibfnamefont{T.}~\bibnamefont{Brandes}},\ }%
  \bibfield{journal}{%
  \Doi{10.1103/PhysRevE.67.066203}{\bibinfo {journal} {Phys. Rev. E}}\ }%
  \textbf{\bibinfo {volume} {67}},\ \bibinfo {pages} {066203} (\bibinfo {year}
  {2003})%
  \bibAnnoteFile{NoStop}{Emary:2003qf}%
\bibitem{Baumann:2010vn}%
  \BibitemOpen
  \bibfield{author}{%
  \bibinfo {author} {\bibfnamefont{K.}~\bibnamefont{Baumann}}, \bibinfo
  {author} {\bibfnamefont{C.}~\bibnamefont{Guerlin}}, \bibinfo {author}
  {\bibfnamefont{F.}~\bibnamefont{Brennecke}},\ and\ \bibinfo {author}
  {\bibfnamefont{T.}~\bibnamefont{Esslinger}},\ }%
  \bibfield{journal}{%
  \bibinfo {journal} {Nature}\ }%
  \textbf{\bibinfo {volume} {464}},\ \bibinfo {pages} {1301} (\bibinfo {year}
  {2010})%
  \bibAnnoteFile{NoStop}{Baumann:2010vn}%
\bibitem{Arecchi:1972kx}%
  \BibitemOpen
  \bibfield{author}{%
  \bibinfo {author} {\bibfnamefont{F.~T.}\ \bibnamefont{Arecchi}}, \bibinfo
  {author} {\bibfnamefont{E.}~\bibnamefont{Courtens}}, \bibinfo {author}
  {\bibfnamefont{R.}~\bibnamefont{Gilmore}},\ and\ \bibinfo {author}
  {\bibfnamefont{H.}~\bibnamefont{Thomas}},\ }%
  \bibfield{journal}{%
  \bibinfo {journal} {Phys. Rev. A}\ }%
  \textbf{\bibinfo {volume} {6}},\ \bibinfo {pages} {2211} (\bibinfo {year}
  {1972})%
  \bibAnnoteFile{NoStop}{Arecchi:1972kx}%
\bibitem{Glauber:1976uq}%
  \BibitemOpen
  \bibfield{author}{%
  \bibinfo {author} {\bibfnamefont{R.~J.}\ \bibnamefont{Glauber}}\ and\
  \bibinfo {author} {\bibfnamefont{F.}~\bibnamefont{Haake}},\ }%
  \bibfield{journal}{%
  \bibinfo {journal} {Phys. Rev. A}\ }%
  \textbf{\bibinfo {volume} {13}},\ \bibinfo {pages} {357} (\bibinfo {year}
  {1976})%
  \bibAnnoteFile{NoStop}{Glauber:1976uq}%
\bibitem{Scully:1994ly}%
  \BibitemOpen
  \bibfield{author}{%
  \bibinfo {author} {\bibfnamefont{M.~O.}\ \bibnamefont{Scully}}\ and\ \bibinfo
  {author} {\bibfnamefont{K.}~\bibnamefont{W\'odkiewicz}},\ }%
  \bibfield{journal}{%
  \bibinfo {journal} {Found. Phys.}\ }%
  \textbf{\bibinfo {volume} {24}},\ \bibinfo {pages} {85} (\bibinfo {year}
  {1994})%
  \bibAnnoteFile{NoStop}{Scully:1994ly}%
\bibitem{Glauber:2007fk}%
  \BibitemOpen
  \bibfield{author}{%
  \bibinfo {author} {\bibfnamefont{R.}~\bibnamefont{Glauber}},\ }%
  \emph{\bibinfo {title} {Quantum Theory of Optical Coherence}}\ (\bibinfo
  {publisher} {Wiley, Weinheim},\ \bibinfo {year} {2007})%
  \bibAnnoteFile{NoStop}{Glauber:2007fk}%
\bibitem{Note2}%
  \BibitemOpen
  \bibinfo {note} {For instance, the frequently employed Wigner function $W$
  has evolution equations with only odd-order derivatives such that quantum
  features start with third order. One then sometimes dispatches all but the
  first-order derivative terms ('truncated Wigner approximation'), thence
  relegating the effects of quantum mechanics to the uncertainty carried by the
  initial distribution. Interestingly, for chaotic dynamics $W$ typically
  develops substructures within Planck cells\cite {Zurek:2001uq}; such
  structures cannot show up in $Q$ which arises from $W$ by averaging over,
  roughly, a Planck cell.}%
  \bibAnnoteFile{Stop}{Note2}%
\bibitem{Note3}%
  \BibitemOpen
  \bibinfo {note} {A. Altland and F. Haake, to be published}%
  \bibAnnoteFile{NoStop}{Note3}%
\bibitem{Chaudhury:2009fk}%
  \BibitemOpen
  \bibfield{author}{%
  \bibinfo {author} {\bibfnamefont{S.}~\bibnamefont{Chaudhury}}, \bibinfo
  {author} {\bibfnamefont{A.}~\bibnamefont{Smith}}, \bibinfo {author}
  {\bibfnamefont{B.~E.}\ \bibnamefont{Anderson}}, \bibinfo {author}
  {\bibfnamefont{S.}~\bibnamefont{Ghose}},\ and\ \bibinfo {author}
  {\bibfnamefont{P.~S.}\ \bibnamefont{Jessen}},\ }%
  \bibfield{journal}{%
  \bibinfo {journal} {Nature}\ }%
  \textbf{\bibinfo {volume} {461}},\ \bibinfo {pages} {768} (\bibinfo {year}
  {2009})%
  \bibAnnoteFile{NoStop}{Chaudhury:2009fk}%
\bibitem{AlLa1}%
  \BibitemOpen
  \bibfield{author}{%
  \bibinfo {author} {\bibfnamefont{I.~L.}\ \bibnamefont{Aleiner}}\ and\
  \bibinfo {author} {\bibfnamefont{A.~I.}\ \bibnamefont{Larkin}},\ }%
  \bibfield{journal}{%
  \bibinfo {journal} {Phys. Rev. {\rm B}}\ }%
  \textbf{\bibinfo {volume} {54}},\ \bibinfo {pages} {14423} (\bibinfo {year}
  {1996})%
  \bibAnnoteFile{NoStop}{AlLa1}%
\bibitem{Gnutzmann:2000uq}%
  \BibitemOpen
  \bibfield{author}{%
  \bibinfo {author} {\bibfnamefont{S.}~\bibnamefont{Gnutzmann}}, \bibinfo
  {author} {\bibfnamefont{F.}~\bibnamefont{Haake}},\ and\ \bibinfo {author}
  {\bibfnamefont{M.}~\bibnamefont{Ku\'s}},\ }%
  \bibfield{journal}{%
  \bibinfo {journal} {Journal of Physics A: Mathematical and General}\ }%
  \textbf{\bibinfo {volume} {33}},\ \bibinfo {pages} {143} (\bibinfo {year}
  {2000})%
  \bibAnnoteFile{NoStop}{Gnutzmann:2000uq}%
\bibitem{Haake}%
  \BibitemOpen
  \bibfield{author}{%
  \bibinfo {author} {\bibfnamefont{F.}~\bibnamefont{Haake}},\ }%
  \emph{\bibinfo {title} {Quantum signatures of chaos, 3rd edition}}\ (\bibinfo
  {publisher} {Springer-Verlag},\ \bibinfo {address} {Berlin},\ \bibinfo {year}
  {2009})%
  \bibAnnoteFile{NoStop}{Haake}%
\bibitem{Trotzky:kx}%
  \BibitemOpen
  \bibfield{author}{%
  \bibinfo {author} {\bibfnamefont{S.}~\bibnamefont{Trotzky}}, \bibinfo
  {author} {\bibfnamefont{Y.-A.}\ \bibnamefont{Chen}}, \bibinfo {author}
  {\bibfnamefont{I.}~\bibnamefont{Flesch}, \bibfnamefont{A.~McCulloch}},
  \bibinfo {author} {\bibfnamefont{U.}~\bibnamefont{Schollw{\"o}ck}}, \bibinfo
  {author} {\bibfnamefont{J.}~\bibnamefont{Eisert}},\ and\ \bibinfo {author}
  {\bibfnamefont{I.}~\bibnamefont{Bloch}},\ }%
  \bibinfo {note} {arXiv:1101.2659}%
  \bibAnnoteFile{NoStop}{Trotzky:kx}%
\bibitem{PhysRevA.79.021608}%
  \BibitemOpen
  \bibfield{author}{%
  \bibinfo {author} {\bibfnamefont{G.}~\bibnamefont{Roux}},\ }%
  \bibfield{journal}{%
  \bibinfo {journal} {Phys. Rev. A}\ }%
  \textbf{\bibinfo {volume} {79}},\ \bibinfo {pages} {021608} (\bibinfo {year}
  {2009})%
  \bibAnnoteFile{NoStop}{PhysRevA.79.021608}%
\bibitem{Zurek:2001uq}%
  \BibitemOpen
  \bibfield{author}{%
  \bibinfo {author} {\bibfnamefont{W.~H.}\ \bibnamefont{Zurek}},\ }%
  \bibfield{journal}{%
  \bibinfo {journal} {Nature}\ }%
  \textbf{\bibinfo {volume} {412}},\ \bibinfo {pages} {712} (\bibinfo {year}
  {2001})%
  \bibAnnoteFile{NoStop}{Zurek:2001uq}%
\end{thebibliography}%
\bibliographystyle{apsrev4-1}

\end{document}